\documentclass[]{aa}
\usepackage{graphicx}
\usepackage{txfonts}
\newcommand{\figcap}[1]{\caption[Another figure]{#1}}
\newcommand{\AJ}[3]{{#1}, AJ, \vol{{#2}}, {#3}.}
\newcommand{\ApJ}[3]{{#1}, ApJ, \vol{{#2}}, {#3}.}

\newcommand{\ApJL}[3]{{#1}, ApJL, \vol{{#2}}, {#3}.}

\newcommand{\AandA}[3]{{#1}, A\&A, \vol{{#2}}, {#3}.}
\newcommand{\AandAsupp}[3]{{#1}, A\&A Supp.\rm, \vol{{#2}}, {#3}.}

\newcommand{\MNRAS}[3]{{#1}, MNRAS\rm, \vol{{#2}}, {#3}.}
\newcommand{\Mess}[3]{{#1}, The Messenger\rm, \vol{{#2}}, {#3}.}

\newcommand{\PASP}[3]{{#1}, PASP\rm, \vol{{#2}}, {#3}.}

\newcommand{\vol}[1]{{\mbox{#1}}}

\newcommand{\meanK}{\mbox{$\langle K\rangle\,$}}

\newcommand{\meanMv}{\mbox{$\langle M_{V}\rangle\,$}}
\newcommand{\meanMk}{\mbox{$\langle M_{K}\rangle\,$}}

\newcommand{\RRab}{RRab\,}
\newcommand{\RRc}{RRc\,}
\newcommand{\RRd}{RRd\,}
\newcommand{\RRe}{RRe\,}

\newcommand{\FeH}{\mbox{[Fe/H]}\,}

\newcommand{\logP}{\mbox{$\log P$}\ }

\newcommand{\mMzero}{\mbox{$(m-M)_0$}\ }

\input{epsf.tex}
\begin{document}
\title{The distance to \object{IC4499} from\\
 $K$-band photometry of 32 RR Lyrae
stars\thanks{Based on data from the European Southern Observatory}}

\author{Jesper~Storm}


\institute{Astrophysikalisches Institut Potsdam,
            An der Sternwarte 16, D-14482 Potsdam, Germany,
	e-mail: jstorm@aip.de
}

\date{Received: September, 2003; Accepted:}

\titlerunning{The distance to IC4499}

\abstract{
  Single epoch $K$-band photometry of 32 RR Lyrae stars in the young galactic
globular cluster \object{IC4499} is presented. The mean magnitudes have been
estimated using the $K$-band template light-curves of Jones, Carney and
Fulbright (\cite{Jones96}). We derive an independent
estimate of the distance and reddening for the cluster based on the new
\meanMk-\logP--\FeH relation from Bono et al.(\cite{Bono03}) which has a
zero-point in agreement with the canonical LMC distance of $(m-M)_0 =
18.5$. For an assumed metallicity of $\FeH=-1.65\pm0.1$ we find
$\mMzero = 16.47 \pm0.04 \pm0.06$ (random and systematic errors
respectively) and $E(B-V)=0.24\pm0.03$ in good agreement
with the findings of Walker and Nemec (\cite{WN96}) based on optical data.
The distance estimate is very robust to uncertainties in the reddening
estimate but naturally depends directly on the adopted zero-point of
the \meanMk-\logP-\FeH relation.

}

\maketitle

\section{Introduction}

  The globular cluster \object{IC4499} (RA: $15^h00^m19^s$, DEC:
$-82^\circ12'50''$; 2000.0)
is among the youngest galactic globular clusters (Ferraro et
al. \cite{Ferraro95}, Fusi Pecci et al., \cite{FusiPecci95}).  It is
of Oosterhoff type I and is thus likely to have zero or even
retrograde rotation around the Galactic center (van den Bergh
\cite{vdBergh93}, Lee and Carney \cite{Lee99}).
It is a fairly low density cluster with
one of the highest numbers of RR Lyrae stars per unit luminosity (Suntzeff,
Kinman and Kraft \cite{Suntzeff91}) containing about 100 RR Lyrae stars.
These extreme characteristics suggests that the cluster
might be a late addition to the Galaxy, which makes the cluster
particularly interesting to study.

Optical color-magnitude diagrams have been
presented by Sarajedini (\cite{Sarajedini93}) and Ferraro et al.
(\cite{Ferraro95}), the latter extending two magnitudes below the
main-sequence turn-off.
The RR Lyrae stars have been the subject of several investigations
(Coutts et al. \cite{Coutts75}, Clements et al. \cite{Clement79},
\cite{Clement86}, Walker and Nemec \cite{WN96}
(WN96 in the following)) since their discovery
(Fourcade and Laborde, \cite{Fourcade69}, and Fourcade et al.,
\cite{Fourcade74}).
These investigations have revealed that the RR Lyrae stars span the
full range of pulsation modes currently known, including stars exhibiting
double mode pulsations, which can be used to determine the mass of these
stars.  WN96 has even suggested that some of the stars might pulsate in the
second overtone. The range of the stellar masses has been shown 
by Sarajedini (\cite{Sarajedini93}) to be 
very limited as judged from the double mode pulsators.

  The reddening towards \object{IC4499}
is fairly high, $E(B - V) = 0.22 \pm 0.02$ (WN96) and thus a
distance estimate which is insensitive to reddening will provide
an important additional constraint on the distance to this cluster.

  The \meanMk - \logP relation for RR Lyrae stars, pioneered by
Longmore et al. (\cite{Longmore86}) and refined further by Longmore
et al. (\cite{Longmore90}), is a very powerful tool for distance
determination. The method is theoretically based on the pulsation equation
from van Albada and Baker (\cite{AB71}) which links the period to the
luminosity of an object.  Bono et al. (\cite{Bono01}), and 
most recently Bono et al. (\cite{Bono03}) (B03 hereinafter)
has refined the method using pulsational stellar models to include a
small but significant metallicity term, and we will use this calibration
in the following.  
It has been shown by Butler (\cite{Butler03}), through a
comparison of inner and outer region RR Lyrae stars in M3, that the method is
insensitive to environmental effects.
It is also rather insensitive to errors in both \FeH and reddening,
and as the amplitudes of RR Lyrae stars in the $K$-band are smaller than
about 0.3~mag it is feasible to apply the relation to single epoch data if
a sufficiently large number of stars can be observed. Using recent ephemerides
for the IC4499 stars we can even phase the observations and use template 
light-curves to improve our estimate of the mean $K$-magnitudes.

\section{The Data}
\label{sec.data}

We used the IRAC-2B near-IR camera (Moorwood et al. \cite{Moorwood92})
mounted at the ESO-MPI 2.2m telescope at La Silla on the night Sep.17/18,
1995.  The detector was a 256 by 256 pixel NICMOS-3 array which was
read out using the so called "non-destructive read" mode. The lens LC,
which provides a projected pixel size of 0.5 arcsec was employed, and
the seeing was typically 1.3 arcsec FWHM.

Two adjacent fields (labeled A and B in the following) including the
center of \object{IC4499} were selected on the basis of the charts in Clement
et al. (\cite{Clement86}) to cover a large number of known RR Lyrae
variables as well as a good range in period.

  The observations were performed in a regular pattern of two dithered
target frames interspersed by two dithered sky frames (several arc-minutes
away) followed by the same sequence again, after offsetting the telescope
by a few arc-seconds. Each frame was made of six 10 seconds integrations which
were averaged on-line by the pre-processing system. For field A, 16 individual
frames were obtained through the $K'$ filter and for field B, 11 such
frames were obtained giving a total of 16 and 11 minutes, respectively,
of on-target integration time. The mean HJD for the two datasets is
2449978.5198 and 2449978.5661 respectively.

\section{Data reduction}

\subsection{Pre-processing}

  All the data were corrected for non-linearity using the 
{\tt ctio.irlincor}
command within IRAF\footnote{IRAF is distributed by the National Optical
Astronomical Observatories, which is operated by AURA, Inc., under cooperative
agreement with the NSF.}. Polynomial coefficients of coeff1=1.0,
coeff2$=3.808477 \times 10^{-2}$, coeff3$=5.37237 \times 10^{-2}$
were adopted on the basis of the reported non-linearity for the array 
(Lidman, \cite{Lidman95}).

  Flatfields were produced as the difference between exposures of the white
screen in the dome with the quartz lights on and off, respectively. 
  The IRAC-2 camera is known to suffer from parasitic light in the flatfields
at  the 5\% level,
and a standard star was mapped in a 4 by 4 grid across the array on the night
of the IC4499 observations as well as on the previous night
in order to determine the necessary correction for the flat fields. The 
counts in each position were fitted by a two dimensional second order
polynomium for each night. The results were in good agreement, to within
3\%, and the averaged correction was applied to the flat field frame. 

  Sky frames were computed as the average of four dithered sky exposures
obtained before and after the science frame. The ``avsigclip'' algorithm of the
IRAF {\tt imcombine} command with a threshold value of $\sigma=5$ was employed
to eliminate stars in the sky fields. For each target
frame the corresponding sky frame was
subtracted and bad pixels were removed by linear interpolation.
The resulting targets frames were finally averaged to produce the
final target frames.

\subsection{Standard star photometry}

Fourteen standard star measurements were performed on the night Oct. 1/2, 1995,
to determine the color terms of the camera in the $J$ and $K'$ filters and the
airmass coefficients. Each star was observed 4 times with the star imaged
in four different positions on the array. The stars were selected from
the list of Carter and Meadows (\cite{Carter95}) to cover as large a
range in $(J -K)$ as possible, i.e. from $-0.02$ to $0.96$, and they were
observed at airmasses ranging from 1.03 to 1.80. Using the conversion
from the Carter system to the CIT system determined by Carter
(\cite{Carter93}) we find:
\begin{eqnarray}
\label{eq.Kcit}
K_{\mbox{\scriptsize CIT}} & = & K'_{\mbox{\scriptsize inst}} 
 + 0.014(\pm0.02) \times (J - K')_{\mbox{\scriptsize inst}} \\
\nonumber                  &   & - 0.17(\pm 0.03)\times X + \mbox{const}
\end{eqnarray}
\noindent
with an RMS scatter of 0.024~mag. We thus find no significant color term,
and will assume a zero effect in the following.

  On the night of the IC4499 observations we also observed 
the standard star HD136879 (Carter and Meadows \cite{Carter95})
before and after the observations
of IC4499 in a pattern of four positions on the detector array. The
airmass for the standard star measurements 
was 1.44 and 2.13 respectively, while the 
airmass for the IC4499 observations ranged from 1.85 to 2.05 (IC4499 is
a very southern object). 
The standard star measurements were interpolated 
in airmass to provide the appropriate offset for the IC4499 data.
Curves of growth were determined from artificial aperture photometry
of the standard star and an aperture radius of 20 pixels (10 arcsec)
was used to determine the offset to the standard system.
From the RMS scatter for Eq.\ref{eq.Kcit} we estimate the error of the
$K$-band zero-point to be 0.025~mag.

\subsection{Photometry}

  The DAOPHOT package (Stetson \cite{Stetson87}) within IRAF was used
to find the stars within the averaged frames and for performing photometry
on them by point spread function (PSF) fitting.
Due to the sparse nature of IC4499 crowding is not an issue for PSF
fitting photometry.

  The aperture corrections for the PSF photometry were determined from a
number of well exposed and reasonably isolated stars in the IC4499 fields.
The aperture photometry was performed after all the stars, apart from
the stars selected for offset determination, had been subtracted by the
IRAF/DAOPHOT {\tt substar} routine. 
The resulting aperture corrections
for five stars in each field agreed to better than 0.02~mag.

The RR Lyrae stars were identified from the finding chart of WN96.
The final magnitudes on the CIT system 
are tabulated in Table \ref{tab.KlogP} together with the periods
and the type of pulsation as determined by WN96.  The pulsation type is
indicated by ``ab'' for fundamental mode pulsators, ``c'' for first overtone,
``d'' for double mode, and ``e?'' for possible second overtone pulsators.
The estimated errors provided by DAOPHOT have been tabulated as well. Note
that the errors on the magnitudes for stars in field B are significantly
larger than in field A due to the shorter integration time.

  The observed magnitudes have been collected at a specific time as
already mentioned in Sect.\ref{sec.data} so with an appropriate ephemeris
for each star we can determine the phase of observation.  Alistair Walker
has kindly provided the full optical light curves from WN96 in machine
readable format and using their period estimates we have determined the
epoch of maximum $V$-light for each of the stars. On this basis we have
determined the phase corresponding to the observed $K$-magnitude. We
have then used the $K$-band template light curves from Jones, Carney
and Fulbright (\cite{Jones96}) to compute the corresponding intensity
averaged mean $K$-magnitude, \meanK, based on the type of pulsation and
the $V$-band amplitude as given by WN96. For the \RRd stars we have
not attempted a phase correction and for the putative \RRe stars we
have assumed that they are actually \RRc stars as done by Castellani,
Caputo and Castellani (\cite{Castellani03}). These estimates of the mean
$K$-magnitude have also been tabulated in Table \ref{tab.KlogP}.

\begin{table}
\caption{\label{tab.KlogP} Periods observed $K$ magnitudes and
the estimated mean $K$ magnitudes
for the RR Lyrae stars in the two fields. The fundamental period is listed for
the \RRd stars. The type of pulsator
as determined by WN96 is listed as well. The last column
indicates in which of the two fields the star was observed.}
\begin{tabular}{r r r r r c c}
\hline\hline 
ID & \multicolumn{1}{c}{\logP} &
\multicolumn{1}{c}{$K$} & \multicolumn{1}{c}{$\sigma (K)$} &
\multicolumn{1}{c}{$\meanK$} &
Type & \multicolumn{1}{c}{Field} \\
\hline
    1 & $-0.2147$ & 15.92 &  0.06 & 15.82 & ab &  A \\
    4 & $-0.2051$ & 15.79 &  0.10 & 15.80 & ab &  B \\
    5 & $-0.2542$ & 15.85 &  0.10 & 15.89 & ab &  B \\
   11 & $-0.1996$ & 15.73 &  0.04 & 15.79 & ab &  A \\
   13 & $-0.2911$ & 16.00 &  0.05 & 16.08 & ab &  A \\
   14 & $-0.3009$ & 15.67 &  0.13 & 15.63 & ab &  B \\
   15 & $-0.2242$ & 15.99 &  0.12 & 16.04 & ab &  B \\
   17 & $-0.3006$ & 16.05 &  0.11 & 15.96 & ab &  B \\
   18 & $-0.3248$ & 16.22 &  0.14 & 16.22 &  d &  B \\
   23 & $-0.2944$ & 16.16 &  0.07 & 16.08 & ab &  A \\
   24 & $-0.2869$ & 15.98 &  0.10 & 16.06 & ab &  B \\
   29 & $-0.4405$ & 16.25 &  0.08 & 16.24 &  c &  A \\
   30 & $-0.2766$ & 15.97 &  0.05 & 15.99 & ab &  A \\
   32 & $-0.4437$ & 15.33 &  0.03 & 15.31 &  c &  A \\
   37 & $-0.2481$ & 15.82 &  0.10 & 15.87 & ab &  B \\
   48 & $-0.2842$ & 16.23 &  0.08 & 16.05 & ab &  A \\
   49 & $-0.3018$ & 15.91 &  0.12 & 15.98 & ab &  B \\
   50 & $-0.2474$ & 15.84 &  0.06 & 15.92 & ab &  A \\
   51 & $-0.3214$ & 16.20 &  0.07 & 16.20 &  d &  A \\
   55 & $-0.4455$ & 16.08 &  0.14 & 16.10 &  c &  B \\
   56 & $-0.5383$ & 16.21 &  0.14 & 16.24 & e? &  B \\
   58 & $-0.3005$ & 15.92 &  0.12 & 15.88 & ab &  B \\
   59 & $-0.3188$ & 15.97 &  0.07 & 15.97 &  d &  A \\
   70 & $-0.2700$ & 15.98 &  0.06 & 15.87 & ab &  A \\
   71 & $-0.3045$ & 15.96 &  0.11 & 15.96 &  d &  B \\
   72 & $-0.1666$ & 15.78 &  0.09 & 15.80 & ab &  B \\
   81 & $-0.4022$ & 15.71 &  0.05 & 15.70 &  c &  A \\
   82 & $-0.2654$ & 15.83 &  0.10 & 15.88 & ab &  B \\
   89 & $-0.4493$ & 16.14 &  0.14 & 16.14 &  c &  B \\
   90 & $-0.3113$ & 16.07 &  0.13 & 16.07 &  d &  B \\
   97 & $-0.5402$ & 16.14 &  0.14 & 16.16 & e? &  B \\
 1575 & $-0.5534$ & 16.44 &  0.09 & 16.41 & e? &  A \\
\hline
 & & mag & mag & mag & & \\
\hline
\end{tabular}
\end{table}

\section{Results}


\subsection{The \meanMk-logP relation}
The \meanMk - \logP relation has been established by Longmore et al.
(\cite{Longmore90}) who determined the slope from photometry of RR Lyrae
stars in 8 different galactic globular clusters. 
Recently, B03 has developed the relation further
to incorporate a small but significant metallicity effect and they find
the relation:
\begin{equation}
\label{eq.logPKZ}
\meanMk = -0.770-2.101\log P + 0.231\FeH
\end{equation}

  For a globular cluster like \object{IC4499} where we assume that there is no
significant spread in \FeH among the RR Lyrae stars, we can simply insert
the best estimate of the metallicity in Eq.\ref{eq.logPKZ} to get a
\meanMk-\logP relation. WN96 discusses the available metallicity
estimates for \object{IC4499} and find a value of $\FeH=-1.65\pm0.1$ based
on spectroscopic measurements of Ca II triplet lines in four RGB stars
by Cannon (\cite{Cannon92}) and from a $(V,(V-I))$ color-magnitude
diagram.

We insert this value in Eq.\ref{eq.logPKZ} and for each star we can then
determine the distance modulus as:
\begin{equation}
\label{eq.mM}
(m-M)_K = \meanK -\meanMk = \meanK - 1.151 - 2.101\log P_0
\end{equation}
The relation refers to the fundamental period and thus it is necessary
to fundamentalize the periods of the higher order pulsators. We adopt
the period ratios as used by WN96, i.e. (P2:P1:P0)=(1.575:1.3439:1.0)
and thus $\log P_0 = \log P_1 + 0.128$ and $\log P_0 = \log P_2 + 0.197$
for the \RRc and \RRe stars, respectively.

\begin{figure*}[htp]
\centering
\epsfxsize=15cm
\epsfbox{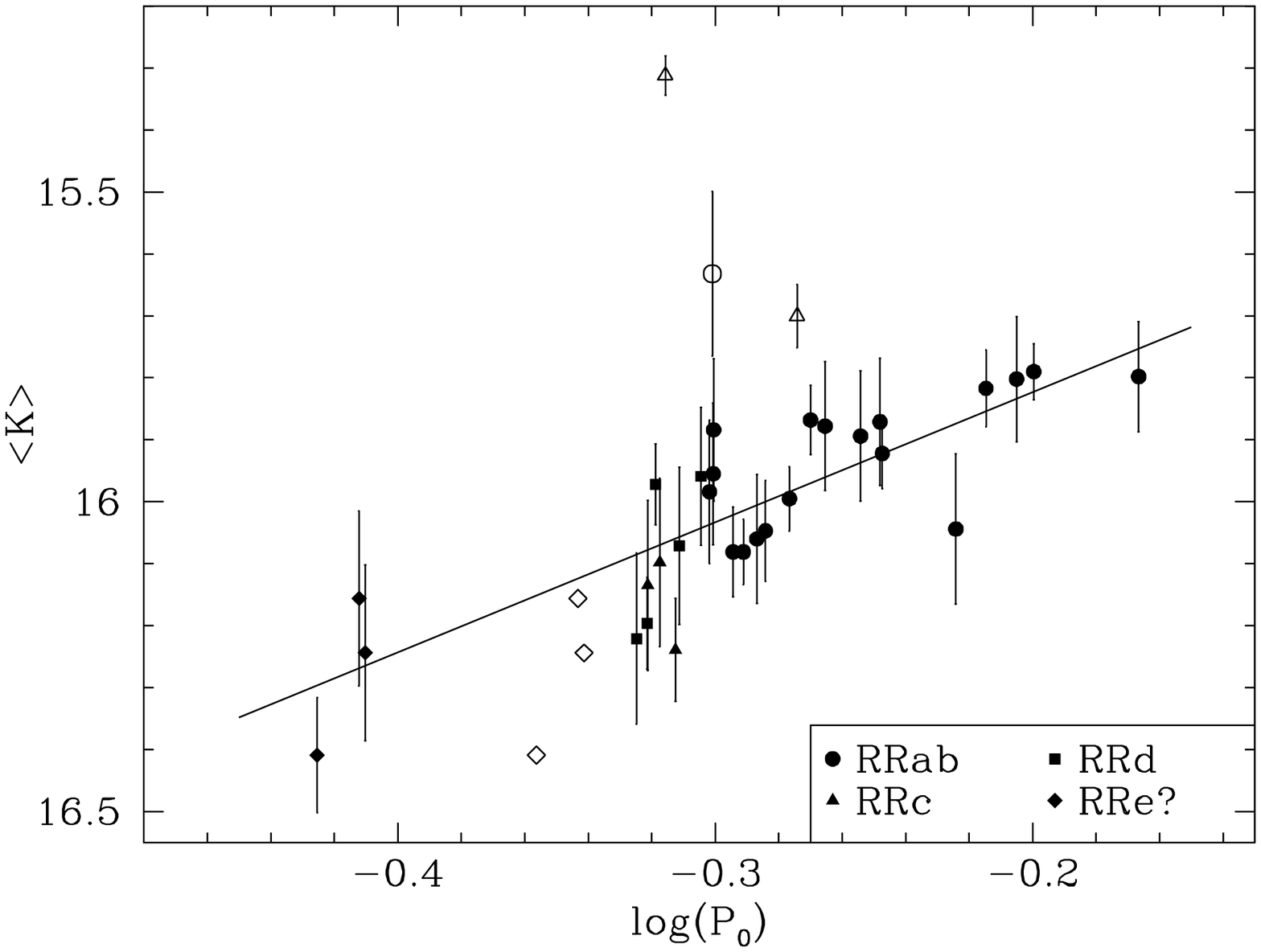}
\figcap{\label{fig.logPK} The $\meanK$ magnitudes plotted against fundamental
period. For the \RRd stars the single epoch $K$ is plotted. The open
symbols were not included in the fit. The filled diamonds labeled
RRe? represents the putative RRe stars from WN96 but fundamentalized as
\RRc stars. The open diamonds show where these stars would appear had
they been fundamentalized as second overtone pulsators.
The line shows Eq.\ref{eq.logPKZ} for $\FeH=-1.65$ and shifted by our 
best distance estimate of $(m-M)_K=16.55$.}
\end{figure*}

In Fig.\ref{fig.logPK} we have plotted the $\meanK$ magnitudes for the
individual stars versus $\log P_0$, the fundamentalized period.
Again we have assumed that the putative \RRe stars are actually \RRc
stars.  They actually fall exactly on the general relation.
If we choose to fundamentalize the period of these stars as second
overtone stars they would appear where the open diamonds in
Fig.\ref{fig.logPK} have been plotted which gives a less satisfying
match.

Inspecting Fig.\ref{fig.logPK} further, we see that three of the stars,
marked with open symbols (V14, V32 and V81), are overly bright
for their period when compared to the bulk of the stars.
WN96 also found both V32 and V81 to be over-luminous in the
optical, and they suspected both stars to have an unresolved red companion. 

The detection limit is fainter than the
faintest RR Lyrae star by about 1~mag in the shallow field (B) so the
sample is complete and can not suffer from Malmquist bias.

\subsection{The reddening and the distance modulus}
  For each star we can now use Eq.\ref{eq.mM} to determine the distance
modulus and we can then compute the weighted mean of the resulting
distances.  If we use the sample of 29 \RRab, \RRc, and \RRd stars
excluding the three outliers we find $(m-M)_K = 16.555\pm0.016$ with
$\sigma=0.087$~mag.  If we include V14 and V32 we find $(m-M)_K =
16.526\pm0.023$ with $\sigma=0.12$~mag, which is only slightly smaller
than the previous result. The three possible \RRe stars do not affect
the distance estimate significantly, independently of how we choose to
fundamentalize them.  Finally, if we choose to simply use the random
phase $K$ magnitudes as a proxy for the mean $K$-magnitude instead of the
template approach described above, we find $(m-M)_K = 16.553 \pm 0.019$
with $\sigma=0.104$, basically an identical result but with a slightly
larger uncertainty.

  Due to the relative insensitivity of the \meanMk-\logP relation
to reddening, we can determine an independent value for the reddening
towards \object{IC4499} from the mean $V$-magnitude of the RR Lyrae
stars as determined by WN96 and the \meanMv-\FeH relation. For consistency
we use the calibrations presented by B03 for both relations,
and thus in addition to Eq. \ref{eq.logPKZ} we adopt
\begin{equation}
\label{eq.MvFeH}
\meanMv = 0.177(\pm 0.07)\FeH + 0.718(\pm 0.07)
\end{equation}
\noindent
which is valid for metal-poor stars ($\FeH<-1.6$).

WN96 find the mean $V$-magnitudes of the RR Lyrae stars to be
$<V_{\mbox{\scriptsize RR}}> = 17.652 \pm 0.006$ and thus $(m-M)_V =
17.226 \pm 0.07$.  Adopting $A_V = 3.1 \times E(B - V)$ and $A_K =
0.11 \times A_V$ (Cardelli et al.  \cite{Cardelli89}) and using the
constraint that the reddening corrected distance estimates from both
the \meanMk-\logP and \meanMv-\FeH relations should be the same, we
can write the relation:
\begin{equation}
(m-M)_V - A_V = (m-M)_K - A_K \\
\end{equation}
from which follows:
\begin{eqnarray}
A_V & = & [(m-M)_V - (m-M)_K]/0.89 \\
A_V & = & 0.755\pm 0.08\\
E(B-V) & = & 0.24\pm0.03
\end{eqnarray}
 
 Our best estimate of the reddening is then $E(B - V) = 0.24 \pm 0.03$ and
the best estimate of the distance to \object{IC4499} is $\mMzero = 16.555 -
0.11\times0.755 = 16.472 \pm 0.016$ where the error estimate
is the formal fitting error only.
The DIRBE reddening map (Schlegel et al. \cite{Schlegel98})
gives $E(B-V)=0.23$ at the position of \object{IC4499}
suggesting that the reddening is completely dominated by galactic
reddening.

In addition to the fitting error we add the error contributions from
the $K$-band zero-point (0.025~mag), from the aperture correction
(0.02~mag), from the uncertainty in $A_K$
(0.009~mag), and from the uncertainty in the metallicity estimate
($0.231\times 0.1=0.023$~mag).  
This gives a final random error of 0.04~mag. 
The zero-point of the PL relation in
Eq.\ref{eq.logPKZ} has an uncertainty of 0.044~mag which should be
considered as a systematic error.  We should also
keep in mind that the metallicity scale is not uniquely defined and that
there might be systematic differences between scales reaching 0.2~dex
(Layden \cite{Layden94}). This would constitute a systematic error of
$0.231\times0.2=0.046$~mag. 
This adds up to an overall systematic error
estimate of 0.06~mag. A much larger potential systematic error
arises from the adopted zero-point of the \meanMk-\logP-\FeH relation.
As an example, Jones et al. (\cite{JCSL}) found a zero-point based
on Baade-Wesselink analysis of field RR Lyrae stars which is about
0.3~mag fainter than the B03 relation.  A similar result was found by
Layden et al. (\cite{Layden96}) from a statistical parallax study of
field RR Lyrae stars.  Walker (\cite{Walker92}) showed that adopting the
Baade-Wesselink and statistical parallax zero-point would cause the LMC
distance estimate from RR Lyrae stars to be about 0.3~mag shorter than the
canonical Cepheid based distance of 18.5~mag. This shows that the 
B03 relation is in good agreement with the canonical Cepheid distance scale 
but it disagrees with the short distance scale.
B03 also show that their relation is
in excellent agreement with the HST parallax measurements of RR~Lyr itself
(Benedict et al. \cite{Benedict02}). 
Kov\'acs \cite{Kovacs03} has redone the Baade-Wesselink analysis for
a sample of galactic Cepheids similar to the one of Jones et al.
(\cite{JCSL}) using the same calibration as has been successfully used
for Cepheids and he now finds agreement with the canonical LMC distance.
We cannot explain the difference between the distance scales with
the current data, but we simply choose to adopt the B03
zero-point to ensure that we are on the canonical Cepheid distance
scale.

  On this basis we find $(m-M)_0 = 16.47\pm0.04$ (random) 
$\pm0.06$ (systematic), or $d=19.7\pm0.4\pm0.6$~kpc,
as our best distance estimate to \object{IC4499}, which 
is in excellent agreement with the value of $(m-M)_0 = 16.45\pm0.05$
found by WN96.
Our reddening value of $E(B-V)=0.24\pm0.03$ also agrees well
with the value of $E(B-V)=0.22\pm0.02$ from WN96.

  We stress that for stellar systems containing a reasonable number of
RR Lyrae stars with known periods and
metallicities, the \meanMk-\logP-\FeH relation combined with single
epoch $K$-band data gives accurate and reddening insensitive distances
for a very modest amount of observational effort as the dominating
source of error is {\em not} the number of individual phase points.

\subsection*{Acknowledgments} The data presented here was acquired during
Directors Discretionary Time, granting of which is gratefully acknowledged.
Thanks is due to Alistair Walker for providing a preprint of
WN96 prior to publication, which helped the identification of the
stars very much. Thanks is also due to Giuseppe Bono
for comments on an early version of this paper and to an anonymous
referee for suggestions which clarified the presentation.


{}
\end{document}